\documentclass[12pt]{article}
\usepackage[english]{babel}
\usepackage{graphicx}
\usepackage{amssymb}
\usepackage{float}
\usepackage{amssymb}
\usepackage{amsmath}
\usepackage{flexisym}
\usepackage{wrapfig}
\usepackage[left=20mm, top=15mm, right=15mm, bottom=15mm, nohead, footskip=10mm]{geometry}
\usepackage{tikz}  

\title{Modified hydrogen recombination rate due to non-Hubble redshift effects}
\author{Krasnov I.V\footnote{iv.krasnov@physics.msu.ru}, Nedelko N.S.\footnote{nedelko@inr.ru}\\\small\emph{Institute for Nuclear Research of Russian Academy of Sciences, 117312 Moscow, Russia}}

\date{(Dated: October 15, 2026)}

\begin{document}

\maketitle

\begin{abstract}
    The accumulation of observational data in recent years has led to a peculiar situation in which measurements of the cosmological parameters made by various collaborations and using different methods yield significantly different values. A solution to this problem may be found by modifying the models used to obtain these values. In this work, we revisit the standard approach to the recombination calculation and investigate a mechanism that can introduce corrections to the rate at which Lyman-alpha photons escape the resonance line. We find that interactions of Lyman-alpha line photons with moving neutral hydrogen atoms can become a source of such a correction. An analysis of CMB data for the modified model favors the existence of a non-Hubble redshift effect and prefers slightly higher values of $H_0$ and smaller values of $\Omega_m$ and $S_8$ compared to the standard $\Lambda$CDM model, as well as a smaller value of the scalar spectrum index $n_s$.
\end{abstract}

\section*{Introduction}
Cosmic microwave background (CMB) measurements provide cosmologists with a glimpse into the early Universe.
The $\Lambda$CDM model is in exceedingly good agreement with the observed data, with many determinations reaching percent-level uncertainties.
Our understanding of the physical conditions in the primordial plasma during the recombination epoch greatly affects the interpretation of observational data.
As we obtain new data, we gain insight into where the inadequacies of our understanding may lie~\cite{CosmoVerseNetwork:2025alb}.
One of the most prominent modern discrepancies in this field is the so-called Hubble tension.
A combination of data from Planck, SPT-3G and ACT under the $\Lambda$CDM assumption provides a CMB-derived value of $H_0 = 67.2 \pm 0.4$ km s$^{-1}$ Mpc$^{-1}$~\cite{SPT-3G:2025bzu}.
The SH0ES Collaboration obtains a higher value, measured by local distance-ladder methods: $H_0 = 73.04 \pm 1.04$ km s$^{-1}$ Mpc$^{-1}$~\cite{Riess:2021jrx}.
There have been many approaches to resolving this tension, including physics beyond the $\Lambda$CDM model, searches for systematics, and more~\cite{CosmoVerseNetwork:2025alb}.
Similarly, there is some tension between different approaches to measuring the amplitude of matter inhomogeneity, with the CMB predicting higher values of the observable $S_8$ parameter than those measured by several (although not all) gravitational lensing surveys~\cite{CosmoVerseNetwork:2025alb}.
More recently, two further issues have appeared. The BAO measurements from the first two data releases of the DESI project are in tension with CMB data within $\Lambda$CDM, which is partially expressed as a preference for a significantly lower total matter density $\Omega_m$~\cite{DESI:2025zgx}. Finally, high-resolution measurements of the CMB from ACT DR6~\cite{ACT:2023kun,AtacamaCosmologyTelescope:2025blo} favor higher $n_s$ values under $\Lambda$CDM, which puts them in moderate tension (worsened when DESI BAO data are added) with simple single-attractor models of inflation~\cite{Ellis:2025zrf}.

In this paper, we approach these problems by modifying the recombination model.
We do not introduce new physics, however, but revisit minor effects that we believe have so far escaped proper attention.
Our main focus is a mechanism by which Lyman-alpha photons leave the resonance due to special-relativistic corrections to the photon energy when such a photon is absorbed by a moving hydrogen atom and subsequently re-emitted.
We study the magnitude of this rate in comparison to the redshift due to the expansion of the Universe, and the way the recombination theory is modified by the introduction of such corrections.
We present the results of a Markov Chain Monte Carlo (MCMC) analysis performed with a modified version of the Boltzmann solver CLASS~\cite{Blas:2011rf} and conclude that our modification augments the standard $\Lambda$CDM model, slightly relaxing the tension between the CMB and DESI BAO~\cite{DESI:2025zgx} in terms of $H_0$ and shifting the value of $n_s$ toward the range preferred by simple attractor models of inflation.

\section{Effects of Lyman-alpha absorption and re-emission}
The initial idea that led us to consider a correction to the recombination rate came from the approach proposed in Refs.~\cite{Shepelev:2024sik,Shepelev:2024guu}.
In these works, it was suggested that special-relativistic corrections to the energy exchange between Lyman-alpha resonance photons and moving hydrogen atoms can make photons leave the resonance line faster than in the standard recombination approach, where this is a result of the redshift induced by the expansion of the Universe.
Note that in our calculations we take a completely different approach to the role this process plays in the recombination epoch compared to the one proposed in those works.

We start with the process $H_0 + \gamma \to H^* \to H_0' + \gamma'$.
The energy of the resulting photon $p_\gamma'$ (in the laboratory reference frame) differs from that of the initial photon $p_\gamma$, possibly removing it from the resonance line.
At a temperature of $5000$ K, we have two important small quantities: $V_H = \sqrt{\frac{3kT}{m_H}} \sim 10^{-5}$, $\frac{h\nu}{m_H} \sim 10^{-8}$.
This means that the main change in the photon energy can be calculated as follows (we obtain this result without approximations in Appendix~\ref{sec:photon_momentum}): 
\begin{equation}
\label{eq:gamma_energy_change}
    \frac{p_\gamma'}{p_\gamma} \approx 1- V_H \left(\cos\alpha(1 - \sin\theta \cos \varphi) + \sin\alpha \sin\theta \sin \varphi\right) -\frac{h\nu}{m_H}(1 - \sin\theta \cos \varphi)+V_H^2(1 - \sin\theta \cos \varphi)
\end{equation}

As shown above, in each correction the random direction of photon emission can play a significant role in determining the resulting photon energy.

One can assume that, since the distribution of the angles $\theta,\varphi$ is isotropic in the rest frame of the excited hydrogen atom, the terms linear in $\sin \varphi$ and $\cos \varphi$ can be omitted on average. In that case, the resulting energy change can be expressed as $\frac{p_\gamma'}{p_\gamma} \approx 1- V_H \cos\alpha -\frac{h\nu}{m_H}+V_H^2$.
These are the Doppler broadening, atomic recoil, and Doppler boosting corrections, which have already been studied in the literature~\cite{Chluba:2008vn}.
Doppler broadening is the strongest of the three corrections and always pushes the photon energy closer to the line center, while the next-order correction, atomic recoil, always leads to a redshift.

Our interest lies in the possibility that the photon energy shifts to the left so much that the photon leaves the resonance line.
We see from Eq.~\eqref{eq:gamma_energy_change} that the randomness of the photon emission direction can actually play a significant role in negating the Doppler broadening.
We start with the simple assumption of a flat resonance spectrum with no wings, i.e., that if the resulting photon has an energy lower than the critical value, then that photon leaves the resonance line:
\begin{equation}
    p_\gamma' < h\nu \frac{\sqrt{1-V_H^2}}{1 + V_H}
\end{equation}
In our notation, this gives the criterion for a photon to leave the resonance line in one $H_0 + \gamma \to H^* \to H_0' + \gamma'$ interaction:
\begin{equation}
\label{eq:sphere_cut}
    V_H (1 + \sin\theta \cos(\varphi+\alpha))-\frac{h\nu}{m_H}(1 - \sin\theta \cos \varphi) < 0
\end{equation}
In Eq.~\eqref{eq:sphere_cut}, the term proportional to $V_H$ cannot take a negative value.
But taking the inequation as a whole, together with the atomic recoil effect, it is possible to satisfy it.
One can notice that, in terms of $\theta,\varphi$, the inequation above poses the problem of finding the area of a normalized sphere cut off by a plane $(1 + x\cos(\alpha) - y \sin(\alpha))-\frac{h\nu}{m_H}\frac{1}{V_H}(1 - x) = 0$. 
That gives us the required geometrical probability for a photon to leave the resonance in one interaction:
\begin{equation}
\label{eq:geometrical_factor}
    P_{\theta,\varphi} = \frac{h\nu}{m_H} \frac{1}{V_H} \frac{1-\cos\alpha}{2}
\end{equation}

\section{Energy loss estimation}
\label{sec:resonance_exit}
If some process causes photons with frequencies near the resonance line to lose energy at a rate $hv_X$, then the time necessary to cross the resonance line from the right border to the left is:
\begin{equation}
    \Delta t_X = \frac{\Delta\nu_\alpha}{v_X}, 
\end{equation}
where $\Delta\nu_\alpha \approx 10^{-4}eV/h$ is the effective Doppler-broadened resonance line width.

The resonance photon energy in the laboratory frame is related to the angle $\alpha$ between the photon momentum and the hydrogen momentum, and to the resonance line frequency: ${h\nu' =h\nu\frac{\sqrt{1-V_H^2}}{1-V_H\cos\alpha}}$.
We still consider the spectrum to be a Doppler-broadened delta function.
For our effect, we can estimate the line exit rate as:
\begin{equation}
    v_{SR} = \int d\nu'  P_{\theta\varphi}(\nu') \frac{n_{1s}}{n} \Gamma \frac{\rho_{\nu'}}{\rho_\nu},
\end{equation}
where $1/\Gamma$ is the mean time between two interactions $1s\xrightarrow{}2p\xrightarrow{}1s$.
During recombination, it is dominated by the average lifetime of a $Ly_\alpha$ photon, $\sim 1sec$.
The factor $\frac{n_{1s}}{n}$ describes the dependence of the interaction probability on the neutral hydrogen density.
The factor $\frac{\rho_{\nu'}}{\rho_\nu} = \left(1 - 3 V_H\cos\alpha\right)$ describes the non-isotropy of the $Ly_\alpha$ photon spectrum in the laboratory reference frame.

Overall, one gets:
\begin{equation}
    v_{SR} = \frac{1}{h}  \frac{n_{1s}}{n} \Gamma \frac{h\nu}{m_H} h\nu \int d(\cos\alpha)  \sqrt{1-V_H^2} \frac{1-\cos\alpha}{2} \frac{1-3V_H\cos\alpha}{(1-V_H\cos\alpha)^2}
\end{equation}
For small values of $V_H$, the integral over the angle $\alpha$ tends to unity.

$\frac{h\nu}{m_H}h\nu \approx \frac{10 eV}{10^9 eV}\times10 eV = 10^{-7} eV$.
Taking, for example, $\Gamma = 0.1 \frac{1}{sec}$, we get the estimate:

\begin{equation}\label{eq:vsr}
    v_{SR} \sim 0.1\times1\times10^{-7}\frac{eV}{sec}\frac{1}{h} \frac{n_{1s}}{n}  = 10^{-8}\frac{eV}{sec}\frac{1}{h} \frac{n_{1s}}{n} 
\end{equation}
and
\begin{equation}
    \Delta t_{SR} \sim \frac{n}{n_{1s}}  \times 10^3sec
\end{equation}

For comparison, $\Delta t_H \sim 10^8$~sec.

Note that our approach differs from the one suggested in Refs.~\cite{Shepelev:2024sik,Shepelev:2024guu}.
The geometrical factor in Eq.~\eqref{eq:geometrical_factor} depends on the angle $\alpha$, and as a result the integral effect is dominated by this factor.
Spectrum modifications give only a minor correction to that value.

But this initial guess has obvious flaws.
First of all, it was shown in the literature that the actual spectrum, which starts as a narrow line, transforms with time into a completely different form~\cite{Chluba:2009gv}.
The most important effect for us is that the line center proper is effectively removed from the resonance because of the higher probability for an excited hydrogen atom to interact with a photon from the Planck spectrum, leading to further excitation and consequent photoionization instead of re-emission of a Lyman-alpha photon.
This not only reduces the overall rate at which neutral hydrogen is excited by photons, but also makes the photons in the wings of the spectral profile contribute more to the process than the photons in the line center.

The effect described in the previous section has the following drawback: in one interaction it can change the photon energy only by a small amount, $\frac{p_\gamma'}{p_\gamma}\sim V_H$ or less, i.e., it is of the order of a Doppler width.
It is a consequence of the wingless resonance line approximation used in that initial estimate.
If we switch from the $\delta$-line approach to a Gaussian-like spectrum, there would be no hard ``borders'' of the resonance line, but rather a steep decline in the probability of interaction in the wings.
The effect described above would still take place, although in a different form.
We can assume that, to effectively remove a photon from the resonance line wings (tens of Doppler widths wide), one interaction would probably be insufficient.
In fact, if a two-interaction shift is needed, that would already mean that the probability obtained in Eq.~\eqref{eq:geometrical_factor} should be squared, suppressing our previous estimate by a factor of the order of $\frac{h\nu}{m_H} \frac{1}{V_H} \sim 10^{-3}$.
As a result, the actual values of $V_{SR}$ can be much smaller than the value obtained in Eq.~\eqref{eq:vsr}.
A proper study of our recombination model with a more realistic spectrum profile will be the focus of another paper.
In this paper, we treat $V_{SR}$ as a new parameter that should have a value somewhere between zero and the initial estimate we obtained in Eq.~\eqref{eq:vsr}.

\section{Modified effective three-level hydrogen atom equations}
The standard approach to the recombination epoch, which is still used in cosmological codes (RECFAST~\cite{Seager:1999bc} and corresponding code in CAMB~\cite{Lewis:1999bs}, CosmoMC~\cite{Lewis:2002ah}, CLASS~\cite{Blas:2011rf}, Cobaya~\cite{Torrado:2020dgo}), is expressed in terms of a so-called three-level system~\cite{Peebles:1968ja,Zeldovich:1969qmq}:

\begin{equation}
\label{eq:redshift_concentration}
    \frac{d}{dt} \left( \frac{n_a}{n}\right) = \frac{\nu_a}{n} \frac{1}{a} \frac{da}{dt} \left[ n_{(\nu^+)} - n_{(\nu^-)} \right] + \frac{R}{n}
\end{equation}
\begin{equation}
\label{eq:N_a_standard}
    \mathcal{N}_{a}  = e^{- \frac{B_1 - B_2}{kT}} + \frac{R a \lambda_a^3}{8\pi \frac{da}{dt} }
\end{equation}
\begin{equation}
\label{eq:R_balance}
    \alpha_c n_e^2 - \beta_c \mathcal{N}_\alpha n_{1s} = R + \Lambda_{2s.1s} \left[ n_{2s} - n_{1s} e^{-(B_1-B_2)/kT} \right]
\end{equation}

Equations~\eqref{eq:N_a_standard} and \eqref{eq:R_balance} give us the expression:
\begin{equation}
\label{eq:N_a_standard2}
    \mathcal{N}_{a}  = \frac{\alpha_c n_e^2 K +(1 + K \Lambda_{2s;1s} n_{1s})e^{-\frac{B_1-B_2}{kT}}}{1 + K (\beta_c + \Lambda_{2s;1s} )n_{1s}}
\end{equation}
where $K = \frac{\lambda_\alpha^3}{8\pi H}$.

The change in the electron density is therefore expressed as:
\begin{equation}
\label{eq:final_dn/dt}
-\frac{d}{dt} \left(\frac{n_e}{n}\right) = \left(\frac{\alpha_c n_e^2}{n} - \beta_c \frac{n_{1s}}{n} e^{-\frac{B_1-B_2}{kT}}\right) \frac{1 + K \Lambda_{2s;1s} n_{1s}}{1 + K (\beta_c + \Lambda_{2s;1s} )n_{1s}}
\end{equation}

We introduce the new effect, which results in a portion of the photons leaving the Lyman-alpha resonance line.
In the Doppler-broadened $\delta$-function approximation, the resonance line spectrum is approximated by a rectangular shape.
Our effect introduces a mechanism by which photons leave the resonance line through the left border, without any balancing effect that would inject photons into the line through the right border.

To account for this effect, we modify Eq.~\eqref{eq:redshift_concentration}, introducing a new parameter $v_\alpha$ describing the additional rate of photons leaving the resonance:
\begin{equation}
    \frac{d}{dt} \left( \frac{n_\alpha}{n}\right) = \frac{\nu_\alpha}{n} \frac{1}{a} \frac{da}{dt} \left[ n_{(\nu^+)} - n_{(\nu^-)} \right]  - \frac{v_{SR}}{n} n_{(\nu^-)}  + \frac{R}{n}
\end{equation}

For simplicity, we introduce a new parameter $V_{SR} = \frac{v_{SR}}{\nu_\alpha}$. 
This leads us to the following modification of Eq.~\eqref{eq:N_a_standard}:
\begin{equation}
    \mathcal{N}_{a}  = e^{- \frac{B_1 - B_2}{kT}} \frac{H}{H + V_{SR}} + \frac{R \lambda_a^3}{8\pi\left( H + V_{SR} \right)}
\end{equation}

Similarly to the standard approach (Eq.~\eqref{eq:N_a_standard2}), we can obtain an expression for $\mathcal{N}_{a}$:
\begin{equation}
    \mathcal{N}_{a}  = \frac{\alpha_c n_e^2 K_{SR} +(\frac{H}{H + V_{SR}} + K_{SR} \Lambda_{2s;1s} n_{1s})e^{-\frac{B_1-B_2}{kT}}}{1 + K_{SR} (\beta_c + \Lambda_{2s;1s} )n_{1s}}
\end{equation}
where $K_{SR} = \frac{\lambda_\alpha^3}{8\pi \left(H + V_{SR}\right)}$.

As a result, the final equation for the electron concentration, Eq.~\eqref{eq:final_dn/dt}, is modified to:
\begin{equation}
\label{eq:change}
-\frac{d}{dt} \left(\frac{n_e}{n}\right) = \left(\frac{\alpha_c n_e^2}{n} - \beta_c \frac{n_{1s}}{n} e^{-\frac{B_1-B_2}{kT}}\left(1-\frac{V_{SR}}{H + V_{SR}}\frac{1}{1 + K_{SR} \Lambda_{2s;1s} n_{1s}}\right)\right) \frac{1 + K_{SR} \Lambda_{2s;1s} n_{1s}}{1 + K_{SR} (\beta_c + \Lambda_{2s;1s} )n_{1s}}
\end{equation}

We consider $V_{SR}$ to be a new model parameter.
Notice that in the case $V_{SR} \to 0$, Eq.~\eqref{eq:change} reduces to the standard form, Eq.~\eqref{eq:final_dn/dt}.
Otherwise, the main difference lies in the coefficient $K_{SR} < K$.
For example, if $\frac{K_{SR}}{K} = \frac{H}{H + V_{SR}} \ll 1$, then the coefficient $C_{SR}=\frac{1 + K_{SR} \Lambda_{2s;1s} n_{1s}}{1 + K_{SR} (\beta_c + \Lambda_{2s;1s} )n_{1s}}$ becomes 1.
The other significant difference lies in the new coefficient $\left(1-\frac{V_{SR}}{H + V_{SR}}\frac{1}{1 + K_{SR} \Lambda_{2s;1s} n_{1s}}\right)$ in front of the expression $\beta_c \frac{n_{1s}}{n} e^{-\frac{B_1-B_2}{kT}}$, which weakens the photoionization rate.
Overall, both effects accelerate recombination.

\section{Numerical results}
First of all, we have checked how significant the corrections that we introduce are.
We have modified the RECFAST~\cite{Seager:1999bc} sections of both CLASS~\cite{Blas:2011rf} and CAMB~\cite{Lewis:1999bs} in accordance with the equations outlined in the previous section, verifying that the resulting CMB spectra are identical in the two codes up to known small differences in numerical treatment (which are also present with standard recombination).
We have also checked that the modified RECFAST at $V_{SR} = 0$ and the unmodified code provide the same result.

\begin{figure}
    \centering
    \includegraphics[width=0.85\linewidth]{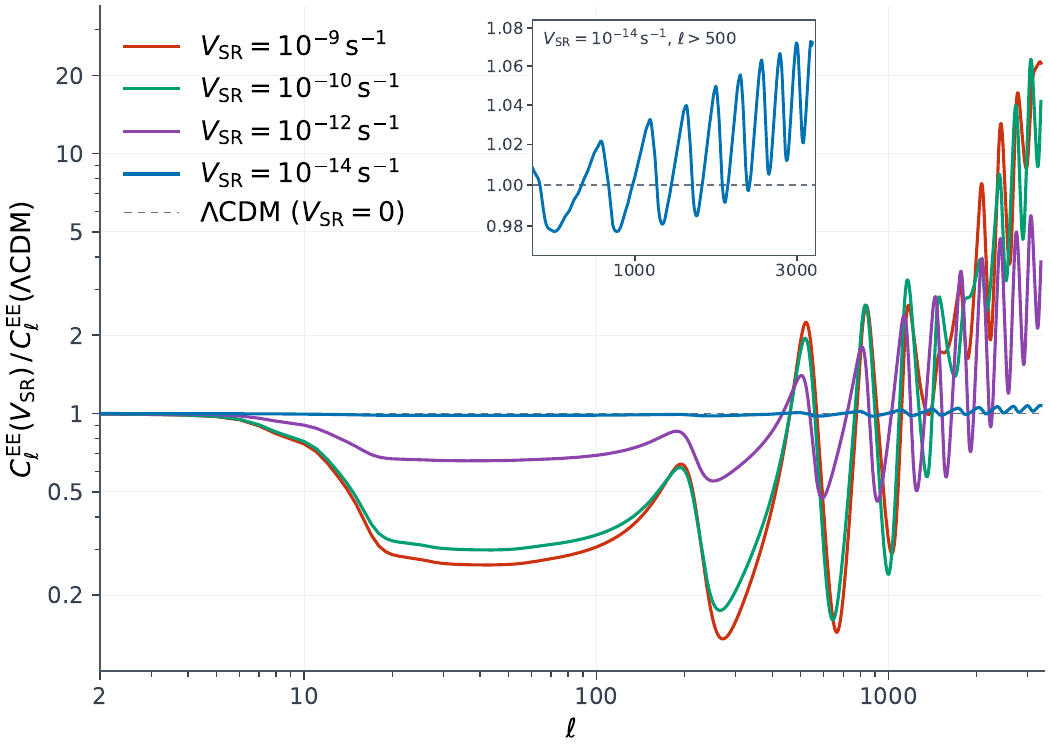}
    \caption{Unlensed \texttt{EE} $C_\ell$ spectra for several reference values of $V_{SR}$ in comparison to the $\Lambda$CDM model.}
    \label{fig:EE}
\end{figure}

\begin{figure}
    \centering
    \includegraphics[width=0.85\linewidth]{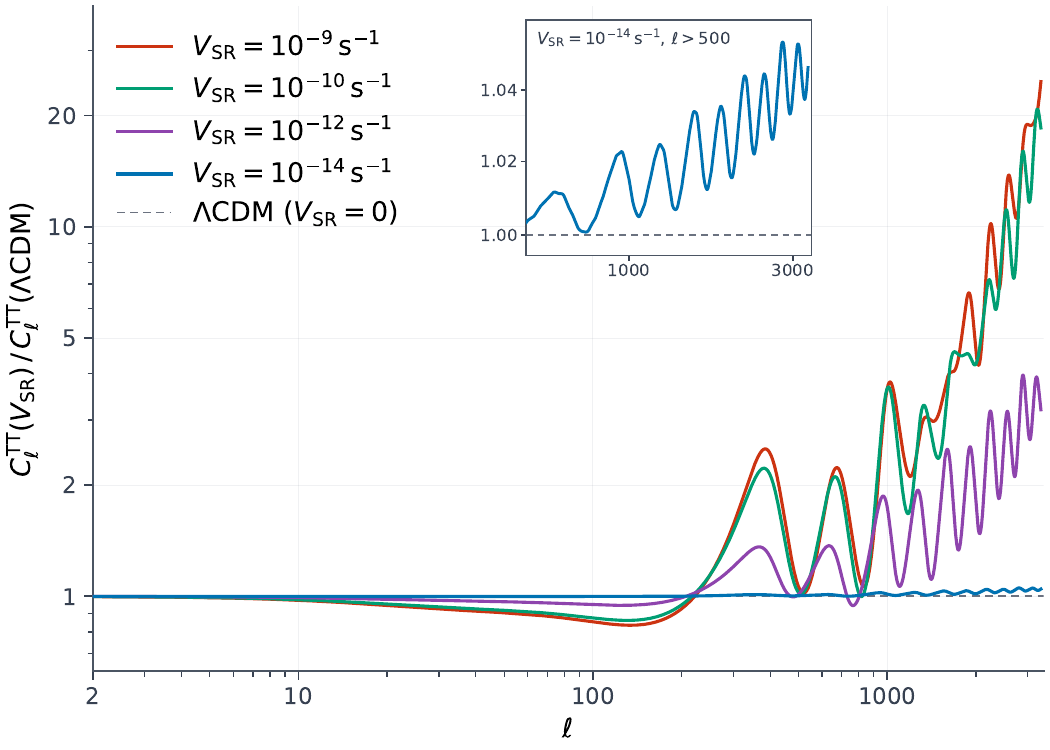}
    \caption{Unlensed \texttt{TT} $C_\ell$ spectra for several reference values of $V_{SR}$ in comparison to the $\Lambda$CDM model.}
    \label{fig:TT}
\end{figure}

\begin{figure}
    \centering
    \includegraphics[width=0.95\linewidth]{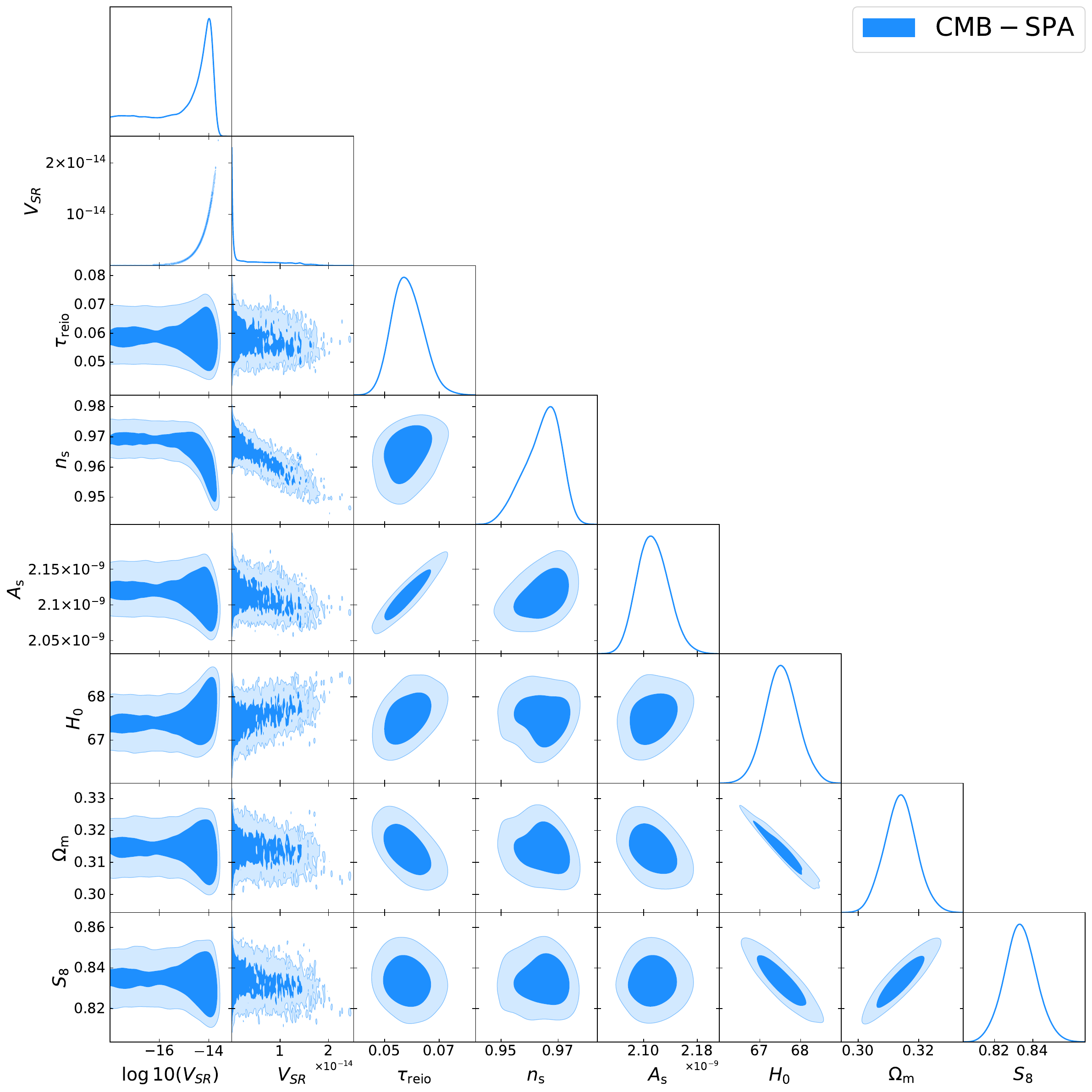}
    \caption{Marginalized constraints on selected cosmological parameters for the CMB-SPA dataset.}
    \label{fig:best_fit}
\end{figure}

We demonstrate the effects of our modification on the unlensed $C_\ell$ spectra in Figs.~\ref{fig:EE} and \ref{fig:TT}, comparing different values of $V_{SR}$ to $\Lambda$CDM for the Planck PR3~\cite{Planck:2018vyg} best-fit cosmology.
Next, we have performed an MCMC analysis using the Cobaya\cite{Torrado:2020dgo} sampler with the modified version of CLASS, letting $V_{SR}$ vary freely~\footnote{Ideally, $V_{SR}$ would be calculated directly from recombination physics. As we cannot yet properly calculate the expected suppression needed to do this, we treat $V_{SR}$ as a sampled parameter in this analysis in order to obtain a statistical estimate of the suppression factor.} (specifically, we sample $\log_{10}V_{SR}$) in addition to the six $\Lambda$CDM parameters chosen as $(\log_{10}A_s,\tau_{reio},H_0,\omega_b,\omega_c)$.
The CMB dataset we use includes:
\begin{itemize}
    \item Planck PR3 $\ell < 30$ TT (Commander)~\cite{Planck:2019nip} and EE (Sroll2)~\cite{Delouis:2019bub} primary CMB spectra
    \item Planck PR3 high-$\ell$ primary TT/TE/EE spectra~\cite{Planck:2019nip} for $\ell_{TT} < 1000, \ell_{TE,EE}<600$ (as in P-ACT) using the \texttt{clipy} version of \texttt{Plik lite}
    \item ACT DR6 primary TT/TE/EE~\cite{ACT:2023kun} \texttt{candl lite}~\cite{Balkenhol:2024sbv} spectra for $600 < \ell < 8500$
    \item SPT-3G D1 primary TT/TE/EE~\cite{SPT-3G:2025bzu} \texttt{candl lite} spectra for $400 < \ell_{TT} < 3000,400 < \ell_{TE,EE} < 4000$
    \item CMB lensing spectra from Planck PR4~\cite{Carron:2022eyg}, ACT DR6~\cite{ACT:2023kun}, and SPT-3G MUSE~\cite{SPT-3G:2024atg}
\end{itemize}
Marginalized posteriors for relevant cosmological parameters are presented in Fig.~\ref{fig:best_fit} and in Table~\ref{tab:placeholder}. 

Cosmologies with a non-zero $V_{SR}$ provide a moderately improved fit compared to $\Lambda$CDM (as evidenced by both the best-fit and the 1D marginalized peak values of $\log_{10}V_{SR}$ being located around $-14$, an $\mathcal{O}(1\%)$ shift of the total $Ly_\alpha$ escape rate).
However, considering that for $\log_{10}V_{SR} < -15$ the cosmology is effectively identical to standard $\Lambda$CDM and that $\Lambda$CDM itself is not heavily disfavored, this leads to an upper limit on $V_{SR}$ of $1.54 \times10^{-14} ~s^{-1}$ at the $95\%$ confidence level.
As a result, the shifts in the parameter constraints are limited, but the effects are nonetheless interesting. 

Most notably, $V_{SR}$ drives down the value of the scalar spectrum index $n_s$ ($0.964\pm0.006$ with a best-fit of $0.960$). 
$V_{SR}$ also induces a slight upward shift in $H_0$ ($67.5\pm0.4$ km s$^{-1}$ Mpc$^{-1}$) with a corresponding downward shift in $\Omega_m$, bringing the CMB constraint somewhat closer to the parameter range preferred by DESI DR2 BAO measurements. Finally, the lower $\Omega_m$ results in a mild decrease in $S_8$ ($0.834\pm0.008$).

\begin{table}
    \centering
    \begin{tabular}{|c|c|c|}
    \hline
       Parameter  & Mean & Best-fit \\
         \hline
    $\log_{10}V_{SR}$ & $-15.1\pm1.3$   & $-14.1$ \\
    $10^{-14}~V_{SR}$   & $<1.55$ & $0.83$ \\
    \hline
    $H_0$ (km s$^{-1}$ Mpc$^{-1}$)     & $67.5\pm0.4$ & $67.6$ \\
    $n_s$     & $0.964\pm0.006$ & $0.960$\\
    $10^9A_s$     & $2.114\pm0.024$ & $2.098$ \\
    $\tau_{reio}$     & $0.058\pm0.006$ & $0.055$ \\
    $10^2\omega_b$     & $2.235\pm0.012$ & $2.230$\\
    $10~\omega_{cdm}$     & $1.202\pm0.010$ & $1.207$\\
    \hline
    $\Omega_m$ & $0.314\pm0.005$     &$0.315$ \\
    $\sigma_8$ & $0.815\pm0.004$     & $0.813$\\
    $S_8$     & $0.834\pm0.008$ & $0.832$ \\
    $r_{drag}$ (Mpc) & $146.94\pm0.29$ & $146.76$ \\
    \hline
    \end{tabular}
    \caption{Cosmological parameter constraints for the CMB-SPA dataset. Two-sided constraints are quoted with $1\sigma$ errors, and the one-sided constraint is cited at the $2\sigma$-equivalent confidence level.}
    \label{tab:placeholder}
\end{table}

\section*{Conclusions}

In this paper, we present a modification of the standard RECFAST recombination model that introduces an additional contribution to the rate at which Lyman-alpha photons escape the resonance line through the left border.
The equations used to determine the electron ionization fraction take a slightly different shape and depend on a new parameter $V_{SR}$, the additional escape rate.
Using a modified version of CLASS we obtain constraints on $V_{SR}$ and on standard cosmological parameters.
We find that using CMB data alone we cannot definitively establish a preference for or against a non-zero $V_{SR}$, constraining it to $V_{SR} <1.54\times10^{-14}s^{-1}$ at the $95\%$ confidence level, while noting that the best-fit region lies at $V_{SR}\sim0.8\times10^{-14}s^{-1}$, an $\mathcal{O}(1\%)$ deviation from the standard $Ly_\alpha$ escape rate.
We provide an example of such a contribution: the ever-present process of Lyman-alpha photon absorption and re-emission by moving neutral hydrogen atoms.
This leads us to conclude that non-Hubble redshift effects that remove Lyman-alpha photons from the resonance line could play a significant role in explaining the observed tensions between cosmological measurements, and therefore should be studied in more detail with extended datasets.

\section*{Acknowledgments}
We would like to thank A.V. Shepelev and D.A. Solovyov for valuable discussions on this topic.

\appendix

\section{Photon momentum calculation}
\label{sec:photon_momentum}

In this appendix, we calculate the change in the photon momentum in a $1s\to2p\to1s$ process.  
We have three input parameters: the hydrogen atom velocity $V_H$, the photon momentum $p_\gamma$, and the angle $\alpha$ between them.
Moreover, these parameters are not independent, as in the hydrogen atom rest frame the photon has to have the $Ly_\alpha$ resonance line frequency $\nu$.
Unless the excited atom interacts further with the medium, for example, by absorbing a thermal photon that excites the hydrogen atom to a higher state, the excited atom emits a photon with almost the same energy as that of the photon it initially absorbed.

\begin{itemize}
    \item [RF №1.] ``Laboratory'' reference frame:
    \begin{equation}
        H : (p_H, 0,0); \quad
        \gamma : (p_\gamma \cos \alpha,  p_\gamma \sin \alpha, 0),
    \end{equation}
    where $p_H = \frac{m_H V_H}{\sqrt{1-V_H^2}};p_\gamma =h\nu \frac{\sqrt{1-V_H^2}}{1 - V_H \cos\alpha}$. Here $h\nu$ is the Lyman-alpha resonance energy.
    \item [RF №2.] Boost to the neutral hydrogen atom rest reference frame:
    \begin{equation}
        H : (0, 0,0); \quad
        \gamma : (h\nu \frac{\cos \alpha - V_H}{1 - V_H \cos\alpha},  h\nu \frac{\sqrt{1-V_H^2} \sin \alpha}{1 - V_H \cos\alpha}, 0).
    \end{equation}
    \item [RF №3.] Perform a rotation:
    \begin{equation}
        H : (0, 0,0); \quad
        \gamma : (h\nu, 0, 0) 
    \end{equation}
    This $xy$ rotation can be described by the angle $\beta$: $\cos \beta =  \frac{\cos\alpha-V_H}{1 - V_H \cos\alpha}, \sin \beta  = \frac{\sqrt{1-V_H^2} \sin \alpha}{1 - V_H \cos\alpha}$.
    
    The excited hydrogen atom has momentum $(p_H^*, 0,0)$, where $p_H^* = \frac{m_H v^*}{\sqrt{1-v^{*^2}}}; v^* = \frac{h\nu}{m_H + h\nu}$.

    \item [RF №4.] Boost to the excited atom rest frame. In this reference frame, we expect that during the spontaneous emission $2p\xrightarrow{}1s$ a photon with energy $h \bar{\nu}$ is emitted isotropically in a random direction:
    \begin{equation}
        H^* : (0, 0,0);\quad \gamma' : (h\bar{\nu} \sin \theta \cos \varphi, h\bar{\nu} \sin \theta \sin \varphi, h\bar{\nu} \cos \theta);
    \end{equation}
    The hydrogen atom has momentum $\vec{p}_{H'} = - \vec{p}_{\gamma'}$, and the photon energy is $h\bar{\nu} =h\nu \frac{1}{\sqrt{1 + 2 \frac{h\nu}{m_H}}}$.

    \item  [RF №3.] Boost to the previous reference frame:
    \begin{equation}
    \gamma' : (\bar{p}_{\gamma_x}, \bar{p}_{\gamma_y}, \bar{p}_{\gamma_z}) = \left(h\nu \frac{h\nu + (m_H + h\nu)\sin \theta \cos \varphi}{m_H + 2 h\nu}, h\nu \frac{\sin \theta \sin \varphi}{\sqrt{1 + 2 \frac{h\nu}{m_H}}}, h\nu \frac{\cos \theta }{\sqrt{1 + 2 \frac{h\nu}{m_H}}}\right)
    \end{equation}
    
    \item  [RF №2.] Rotation in the opposite direction to restore reference frame №2:
    \begin{equation}
    \gamma' : (\bar{p}_{\gamma_x} \cos \beta - \bar{p}_{\gamma_y} \sin \beta, \bar{p}_{\gamma_x} \sin \beta + \bar{p}_{\gamma_y}\cos \beta, \bar{p}_{\gamma_z})
    \end{equation}

    \item  [RF №1.] Boosting to the laboratory reference frame, we obtain the new photon energy:
    \begin{eqnarray}
    \label{hnu_final}
    p_\gamma' &=& \frac{\bar{p_\gamma} + V_H (\bar{p}_{\gamma_x} \cos \beta - \bar{p}_{\gamma_y} \sin \beta)}{\sqrt{{1-V_H^2}}}\nonumber\\   
 &=&  h \nu \Big( \frac{1}{1 + 2 \frac{h\nu}{m_H}}\frac{1}{\sqrt{1-V_H^2}}[1 + \frac{h\nu}{m_H}(1 + \sin\theta \cos \varphi) + \nonumber\\   
    &&\frac{\cos \alpha - V_H}{1- V_H\cos\alpha} V_H (\sin\theta \cos\varphi + \frac{h\nu}{m_H}(1 + \sin\theta \cos \varphi))] - \nonumber\\
    && \frac{1}{\sqrt{1 + 2 \frac{h\nu}{m_H}}}\frac{V_H \sin\alpha}{1 - V_H\cos\alpha}\sin \theta \sin\varphi\Big)
    \end{eqnarray}
    
\end{itemize}

The final change in the photon energy is:
\begin{eqnarray}
\frac{p_\gamma'}{p_\gamma} &=& \frac{1 - V_H\cos\alpha (1 - \sin\theta \cos \varphi) + \frac{h\nu}{m_H} (1 + \sin\theta \cos \varphi) - V_H^2 \sin\theta \cos \varphi - V_H^2 \frac{h\nu}{m_H} (1 + \sin\theta \cos \varphi)}{(1 + 2 \frac{h\nu}{m_H})(1-V_H^2)}\nonumber\\
&& - \frac{V_H \sin\alpha \sin\theta \sin \varphi}{ \sqrt{(1 + 2 \frac{h\nu}{m_H})(1-V_H^2)}}
\end{eqnarray}

\bibliographystyle{utphys}
\bibliography{refs}

\end{document}